\documentclass[11pt]{article}
\usepackage{comment}
\def\nottoobig#1{{\hbox{$\left#1\vcenter to1.111\ht\strutbox{}\right.\n@space$}}}
\newtheorem{theorem}{Theorem}[section]
\newtheorem{corollary}[theorem]{Corollary}
\newtheorem{lemma}[theorem]{Lemma}

\newtheorem{definition}[theorem]{Definition}

\usepackage{epsfig}
\usepackage{amsmath}
\usepackage{amssymb}

\newcommand{\prob}{{\rm Prob}}

\newcommand{\zon}{\zo^n}

\renewcommand{\lll}{\lVert}
\newcommand{\rrr}{\rVert}

\newcommand{\ie}{$\mbox{i.e.}$}
\newcommand{\nat}{\mathbb{N}}
\newcommand{\real}{\mathbb{R}}
\newcommand{\proof}{{\bf Proof.}}

\newcommand{\qedblob}{\mbox{\rule[-1.5pt]{5pt}{10.5pt}}}
\def\literalqed{{\ \nolinebreak\hfill\mbox{\qedblob\quad}}}

\def\qed{\literalqed}

\def\mapping{\rightarrow}
\def\zo{\{0,1\}}
\makeatletter
\def\@listI{\leftmargin\leftmargini \parsep 4.5pt plus 1pt minus 1pt\topsep6pt plus 2pt minus 2pt \itemsep  2pt plus 2pt minus 1pt}

\let\@listi\@listI
\@listi
\makeatother

\author{Marius Zimand\thanks{The author is supported by NSF grant CCF 0634830. Part of this work was done while visiting University of Auckland, New Zealand.}\\ 
Department of Computer and Information Sciences\\
Towson University}
\date{}

\title{Two sources are better than one for increasing the Kolmogorov complexity of infinite sequences}

\begin{document}

\maketitle

%{\LARGE PRELIMINARY DRAFT - DO NOT DISTRIBUTE}

\begin{abstract} 
The  randomness rate of an infinite binary sequence is characterized by the sequence of ratios between the Kolmogorov complexity and the length of the initial segments of the sequence. It is known that there is no uniform effective procedure that transforms one input sequence into another sequence with higher randomness rate. By contrast, we display such a uniform effective procedure having as input two independent sequences with positive but arbitrarily  small constant randomness rate. Moreover the transformation is a truth-table reduction and the output has randomness rate arbitrarily close~to~1. 
\end{abstract}
\section{Introduction}

It is a basic fact that no function can increase the amount of randomness (\ie, entropy) of a finite structure. Formally, if $X$ is a distribution on a finite set $A$, then for any function $f$ mapping $A$ into $A$, the (Shannon) entropy of $f(X)$ cannot be larger than the entropy of $X$. As it is usually the case, the above fact has an analogue in algorithmic information theory: for any finite binary string $x$ and any computable function $f$, $K(f(x)) \leq K(x) + O(1)$, where $K(x)$ is the Kolmogorov complexity of $x$ and the constant depends only on the underlying universal machine. The above inequality has an immediate one-line proof, but the analoguous statement when we move to infinite sequences is not known to hold. For any $\sigma \in [0,1]$, we say that an infinite binary sequence $x$ has randomness rate $\sigma$, if $K(x(1:n))\geq \sigma n$ for all sufficiently large $n$, where $x(1:n)$ denotes the initial segment of $x$ of length $n$.\footnote{The randomness rate of $x$ is very close to the notion of constructive Hausdorff dimension of $x$~\cite{lut:j:dimension, may:j:dimension-kol, rya:j:dimension, sta:j:dimension}; however since this paper is about handling randomness and not about measure-theoretical issues we prefer the randomness terminology.} 
 The question becomes: if $x$ has randomness rate $0 < \sigma < 1$, is there an effective transformation $f$ such that $f(x)$ has randomness rate greater than that of $x$? Unlike the case of finite strings, infinite sequences with positive  randomness rate possess an infinite amount of randomness (even though it is sparsely distributed) and thus it cannot be ruled out that there may be a way to concentrate it and obtain a sequence with higher randomness rate.  

This is a natural question, first raised by Reimann~\cite{rei:t:thesis}, which has received significant attention recently (it is Question 10.1 in the list of open questions of Miller and Nies~\cite{mil-nie:j:open}). So far, there exist several partial results, mostly negative, obtained by restricting the type of transformation.  Reimann and Terwijn~\cite[Th 3.10]{rei:t:thesis}   have shown that for every constant $c < 1$, there exists a sequence $x$ such that if $f$ is a many-one reduction, then the randomness rate of $f(x)$ cannot be larger than $c$. This result has been improved by Nies and Reimann~\cite{nie-rei:c:wtt-Kolm-increase}   to wtt-reductions. More precisely, they showed that for all rational $c \in (0,1)$, there exists a sequence $x$ with randomness rate $c$ such that for all wtt-reductions $f$, $f(x)$ has randomness rate $\leq c$. Bienvenu, Doty, and Stephan~\cite{bie-dot-ste:c:haussdimension}  have obtained an impossibility result for the general case of Turing reductions, which, however, is valid only for \emph{uniform} reductions. Building on the result of Nies and Reimann, they show that for every Turing reduction $f$ and all constants $c_1$ and $c_2$, with $0 < c_1 < c_2 < 1$, there exists $x$ with randomness rate $\geq c_1$ such that $f(x)$, if it exists, has randomness rate $< c_2.$ In other words, loosely speaking, no effective uniform transformation is able to raise the randomness rate from $c_1$ to $c_2$.  Thus the question ``Is there any effective transformation that on input $\sigma \in (0,1]$, $\epsilon > 0$, and $x$, a sequence with randomness rate $\sigma$, produces a string $y$ with randomness rate $\sigma + \epsilon$ ?'' has a negative answer.   On the positive side, Doty~\cite{dot:t:dimextractors}  has shown that for every constant $c$ there exists  a uniform effective transformation $f$ able to transform any $x$ with randomness rate $c \in (0,1]$ into a sequence $f(x)$ that, for infinitely many $n$, has the initial segments of length $n$ with Kolmogorov complexity $\geq (1-\epsilon) n$ (see Doty's paper for the exact statement). However, since Doty's transformation $f$ is a wtt-reduction, it follows from Nies and Reimann's result that $f(x)$ also has infinitely many initial segments with no increase in the Kolmogorov complexity.

In the case of finite strings, as we have observed earlier, there is no effective transformation that increases the absolute amount of Kolmogorov complexity. However, some positive results do exist. Buhrman, Fortnow, Newman, and Vereshchagin~\cite{bfnv:c:increase-kol} show that, for any non-random string of length $n$, one can flip $O(\sqrt{n})$ of its bits and obtain a string with higher Kolmogorov complexity. Fortnow, Hitchcock, Pavan, Vinodchandran, and Wang~\cite{fhpvw:c:extractKol} show that for any $0 < \alpha < \beta < 1$, there is a polynomial-time procedure that on input $x$ with $K(x) > \alpha |x|$, using a constant number of advice bits (which depend on $x$), builds a string $y$ with $K(y) \geq \beta |y|$ and $y$ is shorter than $x$ by only a multiplicative constant. 

Our main result concerns infinite sequences and is a positive one. Recall that Bienvenu, Doty and Stephan have shown that there is no uniform effective way to increase the randomness rate when the input consists of \emph{one} sequence with positive randomness rate.  We show that if instead the input consists of \emph{two} such sequences that are independent, then such a uniform effective transformation exists. 
\begin{theorem} 
\label{t:main}
(Main Result) There exists an effective transformation $f: Q \times \{0,1\}^\infty \times \{0,1\}^\infty \mapping \{0,1\}^\infty$ with the following property: If the input is $\tau \in (0,1]$ and two independent sequences $x$ and $y$ with randomness rate $\tau$, then $f(\tau,x,y)$ has randomness rate $1-\delta$, for all $\delta > 0$. Moreover, the effective transformation is a truth-table reduction.
\end{theorem}
Effective transformations are essentially Turing reductions that are uniform in the parameter $\tau$; see Section~\ref{s:prelim}. Two sequences are independent if they do not contain much common information; see Section~\ref{s:indep}.

One key element of the proof is inspired from Fortnow et al.'s~\cite{fhpvw:c:extractKol}, who showed that a randomness extractor can be used to construct a procedure that increases the Kolmogorov complexity of finite strings. Their procedure for increasing the Kolmogorov complexity runs in polynomial time, but uses a small amount of advice. To obtain the polynomial-time efficiency, they had to use the multi-source extractor of Barak, Impagliazzo, and Wigderson~\cite{bar-imp-wig:c:multisourceext}, which requires a number of sources that is dependent on the initial min-entropy of the sources and on the desired quality of the output. In our case, we are not concerned about the efficiency of the transformation (this of course simplifies our task), but, on the other hand, we want it completely effective (with no advice), we want it to work with just two sources, and we want it to handle infinite sequences.  In place of an extractor, we provide a procedure with similar functionality, using the probabilistic method which is next derandomized in the trivial way by brute force searching. Since we handle infinite sequences, we have to iterate the procedure infinitely many times on finite blocks of the two sources and this necessitates solving some technical issues related to the independence of the blocks.

\section{Preliminaries}
\label{s:prelim}
We work over the binary alphabet $\{0,1\}$. A string is an element of $\{0,1\}^*$ and a sequence is an element of $\{0,1\}^{\infty}$. If $x$ is a string, $|x|$ denotes its length. If $x$ is a string or a sequence and $n,n_1, n_2 \in \nat$, $x(n)$ denotes the $n$-th bit of $x$ and  $x(n_1:n_2)$ is the substring $x(n_1) x(n_1+1) \ldots x(n_2)$. The cardinality of a finite set $A$ is denoted $\lll A \rrr$. Let $M$ be a standard Turing machine. For any string $x$, define the \emph{(plain) Kolmogorov complexity} of $x$ with respect to $M$, as 
\[K_M(x) = \min \{ |p| \mid M(p) = x \}.
\]
 There is a universal Turing machine $U$ such that for every machine $M$ there is a constant $c$ such that for all $x$,
\begin{equation}
\label{e:univ}
K_U(x) \leq K_M(x) + c.
\end{equation}
We fix such a universal machine $U$ and dropping the subscript, we let $K(x)$ denote the Kolmogorov complexity of $x$ with respect to $U$. For the concept of conditional Komogorov complexity, the underlying machine is a Turing machine that in addition to the read/work tape which in the initial state contains the input $p$, has a second tape containing initially a string $y$, which is called the conditioning information. Given such a machine $M$, we define the Kolmogorov complexity of $x$ conditioned by $y$ with respect to $M$ as 
\[K_M(x \mid y) = \min \{ |p| \mid M(p, y) = x \}.
\]
Similarly to the above, there exist  universal machines of this type and they satisfy the relation similar to Equation~\ref{e:univ}, but for conditional complexity. We fix such a universal machine $U$, and dropping the subscript $U$, we let $K(x \mid y)$ denote the Kolmogorov complexity of $x$ conditioned by $y$ with respect to $U$. 

We briefly use the concept of \emph{prefix-free complexity}, which is defined similarly to plain Kolmogorov complexity, the difference being that in the case of prefix-free complexity the domain of the underlying machines is required to be a prefix-free set.

Let $\sigma \in [0,1]$. A sequence $x$ has randomness rate $\sigma$ if $K(x(1:n)) \geq \sigma \cdot n$, for almost every $n$ (\ie, the set of $n$'s violating the inequality is finite).

An effective transformation $f$ is represented by a two-oracle Turing machine $M_f$. The machine $M_f$ has access to two oracles $x$ and $y$, which are binary sequences. When $M_f$ makes the query ``$n$-th bit of first oracle?" (``$n$-th bit of second oracle?"), the machine obtains $x(n)$ (respectively, $y(n)$). On input $(\tau, 1^n)$, where $\tau$ is a rational (given in some canonical representation), $M_f$ outputs one bit. We say that $f(\tau,x,y) = z \in \{0,1\}^\infty$, if for all $n$,  $M_f$ on input $(\tau, 1^n)$ and working with oracles $x$ and $y$ halts and outputs $z(n)$.  (Effective transformations are more commonly called Turing reductions. If $\tau$ would be embedded in the machine $M_f$, instead of being an input, we would say that $z$ is Turing-reducible to $(x,y)$. Our approach emphasizes the fact that we want a family of Turing reductions that is uniform in the parameter $\tau$.)
In case the machine $M_f$ halts on all inputs and with all oracles, we say that $f$ is a truth-table reduction.
\section{Independence}
\label{s:indep}
We need to require that the two inputs $x$ and $y$ that appear in the main result are really distinct, or in the algorithmic-information theoretical terminology, \emph{independent}.

\begin{definition} 
\label{d:indep}
Two infinite binary sequences $x, y$ are independent if for all natural numbers $n$ and $m$,
\[
K(x(1:n) y(1:m)) \geq K(x(1:n)) + K(y(1:m)) - O(\log(n) + \log(m)).
\]
\end{definition}

The definition says that, modulo additive logarithmic terms, there is no shorter way to describe the concatenation of any two initial segments of $x$ and $y$ than having the information that describes the initial segments.

It can be shown that the fact that $x$ and $y$ are independent is equivalent to saying that for every natural numbers $n$ and $m$,
\begin{equation}
K(x(1:n) \mid y(1:m)) \geq K(x(1:n)) - O(\log(n) + \log(m)).
\end{equation}
and
\begin{equation}
K(y(1:m) \mid x(1:n)) \geq K(y(1:m)) - O(\log(n) + \log(m)).
\end{equation}
Thus, if two sequences $x$ and $y$ are independent, no initial segment of one of the sequence can help in getting a shorter description of any initial segment of the other sequence, modulo additive logarithmical terms.

In our main result, the input consists of two sequences $x$ and $y$ that are independent and that have Kolmogorov rate $\sigma$ for some positive constant $\sigma < 1$. We sketch an argument showing that such sequences exist. In  our sketch we take $\sigma = 1/2$.

We start with an arbitrary random (in the Martin-L\"{o}f sense) sequence $x$. Next using the machinery of Martin-L\"{o}f tests relativized with $x$ we infer the existence of a sequence $y$ that is random relative to $x$. From the theory of Martin-L\"{o}f tests, we deduce that there exists a constant $c$ such that for all $m$, $H(y(1:m) \mid x) \geq m  - c$, where $H(\cdot)$ is the prefix-free version of complexity. Since $H(y(1:m)) \leq m + O(\log m)$, for all $m$, we conclude that $H(y(1:m) \mid x ) \geq H(y(1:m)) - O(\log m)$, for all $m$. Therefore,
$H(y(1:m)) \mid x(1:n)) \geq H(y(1:m) \mid x ) - O( \log n) \geq H(y(1:m)) - O(\log n + \log m)$, for all $n$ and $m$. Since the prefix-free complexity $H(\cdot)$ and the plain complexity $K(\cdot)$ are within $O(\log m)$ of each other, it follows that  $K(y(1:m)) \mid x(1:n)) \geq K(y(1:m)) - O(\log n + \log m))$, for all $n$ and $m$. This implies $K(x(1:m) y(1:n) ) \geq K(x(1:n)) + K(y(1:m)) - O(\log(n) + \log(m))$, for all $n,m$. Next we construct $x'$ and $y'$ by inserting in $x$ and respectively $y$, the bit $0$ in all even positions, \ie, $x'= x_1 0 x_2 0 \ldots$ (where $x_i$ is the $i$-th bit of $x$) and $y'= y_1 0 y_2 0 \ldots$.  Clearly, $K(x(1:n))$ and $K(x_1 0 \ldots x_n 0)$ are within a constant of each other, and the same holds for $y$ and $y'$. It follows that $x'$ and $y'$ are independent and have randomness rate $1/2$.

\section{Proof of Main Result}
\subsection{Proof Overview}
We present in a simplified setting the main ideas of the construction. Suppose we have two independent strings $x$ and $y$ of length $n$ such that $K(x) = \sigma n$ and $K(y) = \sigma n$, for some $\sigma > 0$. We want to construct a string $z$ of length $m$ such that $K(z) > (1-\epsilon)m$. The key idea (borrowed from the theory of randomness extractors) is to use a function $E: \zon \times \zon \mapping \zo^m$ such that every large enough rectangle  of $\zon \times \zon$ maps about the same number of pairs  into all elements of $\zo^m$. We say that such a function is \emph{regular} (the formal Definition~\ref{d:regular} has some parameters which quantify the degree of regularity). 
To illustrate the idea, suppose for a moment that we have a function $E: \zon \times \zon \mapping \zo^m$  that, for all subsets $B \subseteq \zon$ with $\lll B \rrr \approx 2^{\sigma n}$,  has the property that any $a \in \zo^m$ has the same number of preimages in $B \times B$, which is of course $\lll B \times B \rrr/2^m$.
Then for any $A \subseteq \zo^m$, $E^{-1}(A) \cap (B \times B)$ has size $\frac{\lll B \times B \rrr}{2^m} \cdot \lll A \rrr$. Let us take $z = E(x,y)$ and let us suppose that $K(z) < (1-\epsilon)m$. Note that the set $B = \{u \in \zo^n \mid K(u) = \sigma n\}$ has size $\approx 2^{\sigma n}$, the set $A = \{v \in \zo^m \mid K(v) < (1-\epsilon)m \}$ has size $< 2^{(1-\epsilon)m}$ and that $x$ and $y$ are in $E^{-1}(A) \cap (B \times B)$. By the above observation the set $E^{-1}(A) \cap (B \times B)$ has size $\leq \frac{2^{\sigma n} \cdot 2^{\sigma n}}{2^{\epsilon m}}$. Since $E^{-1}(A) \cap (B \times B)$ can be enumerated effectively, any pair of strings in $E^{-1}(A) \cap B \times B$ can be described by its rank in a fixed enumeration of $E^{-1}(A) \cap B \times B$. In particular $(x,y)$ is such a pair and therefore $K(xy) \leq 2 \sigma n - \epsilon m$. On the other hand, since $x$ and $y$ are independent, $K(xy) \approx K(x) + K(y) = 2\sigma n$. The contradiction we have reached shows that in fact $K(z) \geq (1-\epsilon)m$.

A function $E$ having the strong regularity requirement stated above may not exist. Fortunately, using the probabilistic method, it can be shown (see Section~\ref{s:regular}) that, for all $m \leq n^{0.99\sigma}$, there exist a function $E: \zon \times \zon \mapping \zo^m$ such that all strings $a \in \zo^m$ have at most $2(\lll B \times B \rrr/2^m)$ preimages in any $B \times B$ as above (instead of $(\lll B \times B \rrr/2^m)$ preimages in the ideal, but not realizable, setting we used above). Once we know that it exists, such a function $E$ can be found effectively by exhaustive search. Then the argument above, with some minor modifications, goes through. In fact, when we apply this idea, we only know that $K(x) \geq \sigma n$ and $K(y) \geq \sigma n$ and therefore we need the function $E$ to satisfy a stronger variant of regularity. However, the main idea remains the same.

Thus there is an effective way to produce a string $z$ with Kolmogorov complexity $(1-\epsilon)m$ from two independent strings $x$ and $y$ of length $n$ and with Kolmogorov complexity $\sigma n$. Recall that, in fact, the input consists of two independent \emph{infinite} sequences $x$ and $y$ with randomness rate $\tau >0$. To take advantage of the procedure sketched above which works for finite strings, we split $x$ and $y$ into finite strings $x_1, x_2, \ldots, x_n, \ldots$, and respectively $y_1, y_2, \ldots, y_n, \ldots$, such that the blocks $x_i$ and $y_i$, of length $n_i$, have still enough Kolmogorov complexity, say $(\tau/2) n_i$, conditioned by the previous blocks $x_1, \ldots , x_{i-1}$ and  $y_1, \ldots , y_{i-1}$. The splitting of $x$ and $y$ into blocks and the properties of the blocks are presented in Section~\ref{s:splitting}. Then using a regular function $E_i : \zo^{n_i} \times \zo^{n_i} \mapping \zo^{m_i}$, we build $z_i = E_i(x_i, y_i)$. By modifying slightly the argument described above, it can be shown that $K(z_i \mid x_1, \ldots, x_{i-1}, y_1, \ldots, y_{i-1}) > (1-\epsilon) m_i$, \ie, $z_i$ has high Kolmogorov complexity even conditioned by the previous blocks $x_1, \ldots , x_{i-1}$ and  $y_1, \ldots , y_{i-1}$. It follows that $K(z_i \mid z_1, \ldots, z_{i-1})$ is also close to $m_i$. We finally take $z = z_1 z_2 \ldots$, and using the above property of each $z_i$, we infer that for every $n$, the prefix of $z$ of length $n$ has randomness rate $> (1-\epsilon)n$. In other words, $z$ has randomness rate $(1-\epsilon)$, as desired.

\subsection{Splitting the two inputs}
\label{s:splitting}
The two input sequences $x$ and $y$ from Theorem~\ref{t:main} are broken into finite blocks $x_1, x_2, \ldots, x_i, \ldots$ and respectively $y_1, y_2, \ldots, y_i, \ldots$. The division is done in such a manner that $x_i$ (respectively, $y_i$) has high Komogorov complexity rate conditioned by the previous blocks $x_1, \ldots, x_{i-1}$ (respectively by the blocks  $y_1, \ldots, y_{i-1}$). The following lemma shows how this division is done.
\begin{lemma} (Splitting lemma)
\label{l:splitting}
Let $x \in \zo^{\infty}$ with randomness rate $\tau$, for some constant $\tau > 0$. Let $0 < \sigma < \tau$. For any $n_0$ sufficiently large, there is $n_1 > n_0$ such that
\[
K(x(n_0+1:n_1) \mid x(1:n_0)) > \sigma(n_1-n_0).
\]
Furthermore, there is an effective procedure that on input $n_0$, $\tau$ and $\sigma$ calculates $n_1$.
\end{lemma}
$\proof$ Let $\sigma'$ be such that $0 < \sigma' < \tau - \sigma$. Take
\[
n_1 = \big \lceil \frac{1-\sigma}{\sigma'} \big \rceil n_0.
\]
Suppose $K(x(n_0+1:n_1) \mid x(1:n_0)) \leq \sigma(n_1-n_0)$. Then $x(1:n_1)$ can be reconstructed from: $x(1:n_0)$, the description of $x(n_0+1:n_1)$ given $x(1:n_0)$, $n_0$, extra constant number of bits describing the procedure.
So
\begin{equation}
\begin{array}{ll}
K(x(1:n_1)) & \leq n_0 + \sigma(n_1 - n_0) + \log n_0 +O(1) \\
& = \sigma n_1 + (1-\sigma)n_0 + \log n_0 +O(1) \\
& \leq \sigma n_1 + \sigma' n_1 + \log n_0 +O(1) \\
& < \tau n_1 \mbox{ (if $n_0$ suffic. large) },

\end{array}
\end{equation}
which is a contradiction if $n_1$ is sufficiently large.~\qed
\smallskip

Now we define the points where we split $x$ and $y$, the two sources.

Take $a$, the point from where the Splitting Lemma holds. For the rest of this section we consider and $b = \big \lceil \frac{1-\sigma}{\sigma'} \big \rceil$.

The following sequence represents the cutting points that will define the blocks. It is defined recursively, as follows: $t_0 = 0$, $t_1 = a$, $t_i = b(t_1 + \ldots + t_{i-1})$. It can be seen that
$t_i = ab(1+b)^{i-2}$, for $i \geq 2$.

Finally, we define the blocks: for each $i \geq 1$, $x_i := x(t_{i-1}+1: t_i)$ and $y_i = y(t_{i-1}+1: t_i)$, and $n_i := |x_i| = |y_i| = ab^2 (1+b)^{i-3}$ (the last equality holds for $i \geq 3$).

We also denote by $\bar{x}_i$ the concatenation of the blocks $x_1, \ldots, x_i$ and by $\bar{y}_i$ the concatenation of the blocks $y_1, \ldots, y_i$.

\begin{lemma}
\label{l:condhigh}
\begin{enumerate}
	\item $K(x_i\mid \bar{x}_{i-1}) > \sigma n_i$, for all $i \geq 2$ (and the analogue relation holds for the $y_i$'s).
	\item $\log|x_i| = \Theta(i)$ and $\log|\bar{x}_i|= \Theta(i)$, for all $i$ (and the analogue relation holds for the $y_i$'s).
\end{enumerate}

\end{lemma}
$\proof$ The first point follows from the Splitting Lemma~\ref{l:splitting}, and the second point follows immediately from the definition of $n_i$ (which is the length of $x_i$) and of $t_i$ (which is the length of $\bar{x}_i$).~\qed

The following facts state some basic algorithmic-information theoretical properties of the blocks $x_1, x_2, \ldots.$ and $y_1, y_2, \ldots$. 

We first recall the following basic fact (for example, see Alexander Shen's lecture notes~\cite{she:t:kolmnotes}).

\begin{theorem}
\label{t:basicineq}
For all finite binary strings $u$ and $v$,
\begin{itemize}
	\item[(i)] $K(vu) \leq K(u) + K(v \mid u) + O(\log K(u) + \log K(v))$.
	\item[(ii)] $K(vu) \geq K(u) + K(v \mid u) -  O(\log K(u) + \log K(v))$.
\end{itemize}
The hidden constants depend only on the universal machine that defines the complexity $K(\cdot)$. 
\end{theorem}
\begin{lemma} 
\label{l:conditionalgen}
For all finite binary strings $u$ and $v$,
\[
\big | K(v \mid u) - \big( K(vu) - K(u) \big) \big| < O(\log |u| + \log |v|).
\]
\end{lemma}
$\proof$ Theorem~\ref{t:basicineq} implies
$\big | K(v \mid u) - \big( K(vu) - K(u) \big) \big| < O(\log K(u) + \log K(v))$.
Since $K(u) \leq |u| + O(1)$ and $K(v) \leq |v| + O(1)$, the conclusion follows.~\qed

\begin{lemma} 
\label{l:concat}
For all $i$ and $j$,

\begin{itemize}
	\item[(a)] $\big| K(\bar{y}_i \bar{x}_j) - \big( K(\bar{y}_i) + K(\bar{x}_j) \big) \big| < O(i+j)$.
	\item[(b)] $\big| K(\bar{x}_i \bar{y}_j) - \big( K(\bar{x}_i) + K(\bar{y}_j) \big) \big| < O(i+j)$.
\end{itemize}
\end{lemma}
$\proof$  We prove $(a)$ ($(b)$ is similar). 
\begin{equation}
\label{e:1}
\begin{array}{ll}
K(\bar{y}_i \bar{x}_j) & \leq K(\bar{y}_i) + K(\bar{x}_j) + O(\log (K(\bar{y}_i) + \log (K(\bar{x}_j)) \\
& \leq K(\bar{y_i}) + K(\bar{x}_j) + O(\log |\bar{y}_i|  + \log |\bar{x}_j|) \\
& =  K(\bar{y}_i) + K(\bar{x}_j) + O(i+j).
\end{array}
\end{equation}
The first line follows from Theorem~\ref{t:basicineq} (i) (keeping in mind that $K(v|u) \leq K(v) + O(1)$). For the last line we took into account that $\log |\bar{x}_j| = O(j)$  and $\log |\bar{y}_i| = O(i)$.

On the other hand,
\begin{equation}
\label{e:2}
\begin{array}{ll}
K(\bar{y}_i \bar{x}_j) & \geq K(\bar{y}_i) + K(\bar{x}_j) - O(\log |\bar{y}_i|  + \log |\bar{x}_j|) \\
& = K(\bar{y}_i) + K(\bar{x}_j) - O(i+j).
\end{array}
\end{equation}
The first line follows from the independence of $x$ and $y$.

Combining equations~(\ref{e:1}) and~(\ref{e:2}), the conclusion follows.~\qed

\begin{lemma} 
\label{l:conditional}
For all $i$ and $j$,
\begin{itemize}
	\item[(a)] $\big| K(x_i \mid \bar{x}_{i-1}\bar{y}_j) - K(x_i \mid \bar{x}_{i-1}) \big| < O(i+j)$.
	\item[(b)] $\big| K(y_i \mid \bar{x}_j \bar{y}_{i-1}) - K(y_i \mid \bar{y}_{i-1}) \big| < O(i+j)$.
\end{itemize}
\end{lemma}
$\proof$ We prove $(a)$ ($(b)$ is similar). We first evaluate $K(x_i \mid \bar{x}_{i-1}\bar{y}_j )$.
From Lemma~\ref{l:conditionalgen},
\begin{equation}
\label{e:3}
\big| K(x_i \mid \bar{x}_{i-1}\bar{y}_j ) - \big( K(x_i \bar{x}_{i-1} \bar{y}_{j})  - K(\bar{x}_{i-1}\bar{y}_{j})\big) \big| < O(i+j).
\end{equation}
It is easy to check that $K(x_i \bar{x}_{i-1} \bar{y}_{j})$ is within $O(i)$ from  $K(\bar{x}_{i-1} x_{i} \bar{y}_{j})$. Thus, we can substitute $K(x_i \bar{x}_{i-1} \bar{y}_{j})$ by $K(\bar{x}_{i} \bar{y}_{j})$ and obtain,
\begin{equation}
\label{e:4}
\big| K(x_i \mid \bar{x}_{i-1}\bar{y}_j ) - \big( K(\bar{x}_{i} \bar{y}_{j})  - K(\bar{x}_{i-1}\bar{y}_{j}) \big) \big| < O(i+j).
\end{equation}
Next, by Lemma~\ref{l:concat}, $ \big| K(\bar{x}_{i} \bar{y}_{j})- \big( K(\bar{x}_i) + K(\bar{y}_{j}) \big) \big| < O(i+j)$ and $ \big |K(\bar{x}_{i-1} \bar{y}_{j}) - \big( K(\bar{x}_{i-1}) + K(\bar{y}_{j}) \big) \big| < O(i+j)$. Plugging these inequalities in Equation~(\ref{e:4}), we get 
\begin{equation}
\label{e:5}
\big| K(x_i \mid \bar{x}_{i-1}\bar{y}_j ) - \big( K(\bar{x}_i) - K(\bar{x}_{i-1}) \big) \big| < O(i+j).
\end{equation}
We next evaluate $K(x_i \mid \bar{x}_{i-1})$. From Lemma~\ref{l:conditionalgen},
\begin{equation}
\label{e:6}
\big| K(x_i \mid \bar{x}_{i-1} ) - \big( K(x_i \bar{x}_{i-1} )  - K(\bar{x}_{i-1})\big) \big| < O(i+j).
\end{equation}
Using the inequality $\big| K(x_i \bar{x}_{i-1} ) - K(\bar{x}_{i-1} x_{i} ) \big| < O(1)$, we obtain
\begin{equation}
\label{e:7}
\big| K(x_i \mid \bar{x}_{i-1}) - \big( K(\bar{x}_i) - K(\bar{x}_{i-1}) \big) \big| < O(i+j).
\end{equation}
From Equations~(\ref{e:5}) and~(\ref{e:7}), the conclusion follows.~\qed
\begin{lemma}
\label{l:combination}
For all $i$, $K(x_i y_i \mid \bar{x}_{i-1} \bar{y}_{i-1}) \geq K(x_i \mid \bar{x}_{i-1} \bar{y}_{i-1}) + K(y_i  \mid \bar{x}_{i-1} \bar{y}_{i-1}) - O(i)$.
\end{lemma}
$\proof$ The conditional version of the inequality in Theorem~\ref{t:basicineq} holds true, \ie, for all strings $u, v$ and $w$, $K(uv \mid w) \geq K(u \mid w) + K(v \mid uw) - O(\log K(u) + \log K(v))$. Thus, keeping into account that $K(x_i) \leq |x_i| + O(1) = 2^{O(i)}$ and $K(y_i) \leq |y_i| + O(1) = 2^{O(i)}$, we get
\[
K(x_i y_i \mid \bar{x}_{i-1}  \bar{y}_{i-1}) \geq K(x_i \mid \bar{x}_{i-1}  \bar{y}_{i-1}) + K(y_i \mid x_i \bar{x}_{i-1}  \bar{y}_{i-1}) - O(i).
\]
Note that $K(y_i \mid x_i \bar{x}_{i-1}  \bar{y}_{i-1})$ and $K(y_i \mid \bar{x}_{i}  \bar{y}_{i-1})$ are within a constant of each other, and therefore
\[
K(x_i y_i \mid \bar{x}_{i-1}  \bar{y}_{i-1}) \geq K(x_i \mid \bar{x}_{i-1}  \bar{y}_{i-1}) + K(y_i \mid  \bar{x}_{i}  \bar{y}_{i-1}) - O(i).
\]
Next, we note that $ K(y_i \mid  \bar{x}_{i}  \bar{y}_{i-1}) \geq  K(y_i \mid  \bar{y}_{i-1}) - O(i) \geq
 K(y_i \mid  \bar{x}_{i-1}  \bar{y}_{i-1})  - O(i)$, where the first inequality is derived from Lemma~\ref{l:conditional}. The conclusion follows.~\qed

\subsection{Regular functions}
\label{s:regular}
The construction of $z$ from $x$ and $y$ proceeds block-wise: we take as inputs the blocks $x_i$ and $y_i$ and, from them, we build $z_i$, the $i$-th block of $z$. The input strings $x_i$ and $y_i$, both of length $n_i$, have Kolmogorov complexity $\sigma n_i$, for some positive constant $\sigma$, and the goal is to produce $z_i$, of length $m_i$ (which will be specified later), with Kolmogorov complexity $(1-\epsilon)m_i$, for positive $\epsilon$ arbitrarily small. This resembles the functionality of randomness extractors and, indeed, the following definition captures a property similar to that of extractors that is sufficient for our purposes.

\begin{comment}
\begin{definition}  The min-entropy of a distribution $X$ on $\zo^n$ is $\min_{a \in \zo^n} (\log \frac{1}{\prob (X=a)})$. If a distribution $X$ on $\zo^n$ has min-entropy $k$ then for all $a \in \zo^n$, $\prob(X=a) \leq 2^{-k}$.
\end{definition}

\begin{definition} The statistical distance between two distributions $X$ and $Y$ over the same finite probabilistic space is $\Delta(X,Y) = \frac{1}{2} (\sum_a |\prob(X = a) - \prob(Y=a)|)$, where the sum is over all elements $a$ of the probabilistic space.
\end{definition}

\begin{definition}
A function $E: \zo^n \times \zo^n \mapping \zo^m$ is a $(k,\epsilon)$ two-source extractor if for any two distributions $X_1$ and $X_2$ over $\zo^n$ with min-entropy $k$, $E(X_1, X_2)$ is $\epsilon$-close to the unifrom distribution on $\zo^m$.
\end{definition}
\end{comment}
\begin{definition}
\label{d:regular}
A function $f: \zo^n \times \zo^n \mapping \zo^m$ is $(\sigma, c)$-regular, if for any $k_1, k_2 \geq \sigma n$, any two subsets $B_1 \subseteq \zo^n$ and $B_2 \subseteq \zo^n$ with $\lll B_1 \rrr = 2^{k_1}$ and $\lll B_2 \rrr = 2^{k_2}$ have the following property: for any $a \in \zo^m$, 
\[
\lll f^{-1}(a) \cap (B_1 \times B_2) \rrr \leq \frac{c}{2^m} \lll B_1 \times B_2 \rrr.
\]
\end{definition}

Remarks: Let $[N]$ be the set $\{1, \ldots, N\}$. We identify in the standard way $\zo^n$ with $[N]$, where $N = 2^n$. We can view $[N] \times [N]$ as a table with $N$ rows and $N$ columns and a function  $f: [N] \times [N] \mapping [M]$ as an assignment  of a color chosen from $[M]$ to each cell of the table. The function $f$ is $(\sigma, c)$-regular if in any rectangle of size $[K] \times [K]$, with $k \geq \sigma n$, no color appears more than a fraction of $c/M$ times. (The notion of regularity is interesting for small values of $c$ because it says that in all rectangles, unless they are small, all the colors appear approximately the same number of times; note that if $c = 1$, then all the colors appear the same number of times.)

We show using the probabilistic method that for  any $\sigma > 0$, $(\sigma, 2)$-regular functions exist. Since the regularity property for a function $f$ (given via its truth table) can be effectively tested, we can effectively construct $(\sigma, 2)$- regular functions by exhaustive search

We take $f: [N] \times [N] \mapping [M]$, a random function. First we show that with positive probability such a function satisfies the definition of regularity for sets $A$ and $B$ having size $2^k$, where $k$ is \emph{exactly} $\lceil \sigma n\rceil$. Let's temporarily call this property the \emph{weak regularity} property. We will  show that in fact weak regularity  implies the regularity property as defined above (\ie, the regularity should hold for all sets $B_1$ and $B_2$ of size $2^{k_1}$ and respectively $2^{k_2}$, for $k_1$ and $k_2$ \emph{greater or equal} $\lceil \sigma n\rceil$).

\begin{lemma}
For every $\sigma > 0$, if $M \leq N^{0.99 \sigma}$, then it holds with probability $> 0$ that $f$ satisfies the $(\sigma, 2)$- weak regularity property as defined above.
\end{lemma}

$\proof$. 

Fix $B_1 \subseteq [N]$ with $\lll B_1 \rrr = N^\sigma$ (to keep the notation simple, we ignore truncation issues).

Fix $B_2 \subseteq [N]$ with $\lll B_2 \rrr = N^\sigma$.

Let $j_1 \in B_1 \times B_2$ and $j_2 \in [M]$ be fixed values. As discussed above, we view $[N] \times [N]$ as a table with $N$ rows and $N$ columns. Then  $B_1 \times B_2$ is a rectangle in the table, $j_1$ is a cell in the rectangle, and $j_2$ is a color out of $M$ possible colors.

Clearly, $\prob(f(j_1) = j_2) = 1/M$.

%Expected number of cells in the rectangle $A \times B$ that are colored $j_2$ is $\mu = %\frac{N^\sigma \cdot N^\sigma}{M}$.

By Chernoff bounds,
\begin{equation*}
\prob \bigg( \bigg ( \frac{\mbox{no. of $j_2$-colored cells in $B_1 \times B_2$}}{N^\sigma \cdot N^\sigma} - \frac{1}{M}\bigg )  > \frac{1}{M} \bigg ) <  e^{-(1/M) \cdot N^\sigma \cdot N^\sigma \cdot(1/3)}.
\end{equation*}
By the union bound
\begin{equation}
\label{e:eq1}
\prob( \mbox{ the above holds for some $j_2$ in $[M]$ } ) < M e^{-(1/M) \cdot N^\sigma \cdot N^\sigma \cdot(1/3)}.
\end{equation}
 The number of rectangles $B_1 \times B_2$ is 
 \begin{equation}
 \label{e:eq2}
 \begin{array}{ll}
  {N \choose N^\sigma} \cdot {N \choose N^\sigma} \leq \bigg ( \big(\frac{eN}{N^\sigma} \big)^{N^\sigma} \bigg)^2 
 = e^{2N^{\sigma}} \cdot e^{2N^\sigma \cdot (1-\sigma) \ln N}.
\end{array}
\end{equation}
%Thus the probability that there is a rectangle $A \times B$ and a value $j_2$ where the bias %is $>  \frac{1}{M}$ is bounded by (\ref{e:eq1}) $\cdot $ (\ref{e:eq2} ).

Note that if there is no rectangle $B_1 \times B_2$ and $j_2$ as above, then $f$ satisfies the weaker $(\sigma, 2)$-regularity property.

Therefore we need that the product of the right hand sides in equations~(\ref{e:eq1}) and~ (\ref{e:eq2} ) is $< 1$.

This is equivalent to
\[
(1/M) \cdot N^{2\sigma} \cdot{1/3} - \ln(M) > 2N^\sigma +2N^\sigma \cdot (1-\sigma)\ln N,
\]
which holds true for $M \leq N^{0.99 \sigma}$.~\qed
\begin{comment}
{\bf Note.} Actually we need something weaker than extractor. We need only the following:
\smallskip

For any $\tau \in (0,1)$ for any $n$ sufficiently large, and any $m < \tau n$, there is a function

$E: \zo^n \times \zo^n \mapping \zo^m$ 

such that for every $a \in \zo^m$,  for any $k \geq \tau n$, for any $A \subseteq \zo^n$ with $\lll A \rrr \geq 2^k$ and for any $B \subseteq \zo^n$ with $\lll B \rrr \geq 2^k$, it holds that
\[
\lll E^{-1}(a) \cap A \times B \rrr \leq \frac{2}{2^m} \lll A \times B \rrr.
\]
\smallskip

This can be shown as in the Lemma, but it's enough to take $\epsilon = 1$. We use only the Chernoff bound
\[
\prob( 1/n \sum_{i=1}^n X_i > (1+\epsilon)p) < e^{-\epsilon^2 p n /3}.
\]
\end{comment}

As promised, we show next that weak regularity implies regularity.

\begin{lemma}Let $f: \zo^n \times \zo^n \mapping \zo^m$ such that for every $B_1 \subseteq \zo^n$ with $\lll B_1 \rrr = 2^k$, for every $B_2 \subseteq \zo^n$ with $\lll B_2 \rrr = 2^k$, and for every $a \in \zo^m$ it holds that
\[
\lll f^{-1}(a) \cap (B_1 \times B_2) \rrr \leq p.
%\prob_{x \in A, y \in B} (f(x,y)=a) \leq p.
\]
Then for every $k_1 \geq k$ and every $k_2 \geq k$, for every $B_1' \subseteq \zo^n$ with $\lll B_1' \rrr = 2^{k_1}$, for every $B_2' \subseteq \zo^n$ with $\lll B_2' \rrr = 2^{k_2}$, and for every $a \in \zo^m$ it holds that
\[
\lll f^{-1}(a) \cap (B_1' \times B_2') \rrr \leq p.
%\prob_{x \in A', y \in B'} (f(x,y)=a) \leq p.
\]
\end{lemma}
$\proof$. We partition $B_1'$ and $B_2'$ into subsets of size $2^k$. So, $B_1' = A_1 \cup A_2 \cup \ldots \cup A_s$, with $\lll A_i \rrr =2^k$, $i=1, \ldots, s$ and $B_2 = C_1 \cup C_2 \cup \ldots \cup C_t$, with $\lll C_j \rrr =2^k$, $j=1, \ldots, t$.
Then,
\begin{equation*}
\begin{array}{ll}
\lll f^{-1}(a) \cap (B_1' \times B_2') \rrr & = \sum_{i=1}^{s} \sum_{j=1}^t \lll f^{-1}(a) \cap (A_i \times C_j) \rrr \\
&\leq \sum_{i=1}^{s} \sum_{j=1}^t p \cdot \lll A_i \times C_j \rrr \\
&= p \cdot \lll B_1' \times B_2' \rrr.
%\prob_{x \in A', y \in B'} (f(x,y) = a) & = \sum_{i=1}^{s} \sum_{j=1}^t \prob_{x \in A', y \in B'}(f(x,y) = a %\mbox{ and } x \in A_i \mbox{ and } y \in B_j) \\
%& =  \sum_{i=1}^{s} \sum_{j=1}^t \prob_{x \in A', y \in B'} (x \in A_i \mbox{ and } y \in B_j) \\
%& \hspace{3cm} \cdot \prob_{x \in A', y \in B'} (f(x,y) = a \mid x \in A_i \mbox{ and } y \in B_j) \\
%& \leq p \cdot \sum_{i=1}^{s} \sum_{j=1}^t \prob_{x \in A', y \in B'} (x \in A_i \mbox{ and } y \in B_j) = p.
\end{array}
\end{equation*}

\subsection{Increasing the randomness rate}
We proceed to the proof of our main result, Theorem~\ref{t:main}.

We give a ``global" description of the effective mapping $f : Q \times \zo^\infty \times \zo^\infty \mapping \zo^\infty$. It will be clear how to obtain the $n$-th bit of the output in finitely many steps, as it is formally required. 
\medskip

{\bf Construction}
\smallskip

\fbox{
\vbox{

\emph{Input:} $\tau \in Q \cap (0,1]$, $x, y \in \zo^\infty$ (the sequences $x$ and $y$ are oracles to which the procedure has access).

\emph{Step 1:} Split $x$ into $x_1, x_2, \ldots, x_i, \ldots$ and  split  $y$ into $y_1, y_2, \ldots, y_i, \ldots $, as described in Section~\ref{s:splitting} taking $\sigma = \tau/2$ and $\sigma' = \tau/4$.

For each $i$, let $|x_i| = |y_i| = n_i$ (as described in Section~\ref{s:splitting}).

By Lemma~\ref{l:condhigh}, $K (x_i \mid \bar{x}_{i-1}) > \sigma n_i$ and $K^x(y_i \mid \bar{y}_{i-1}) > \sigma n_i$.

\emph{Step 2:} As discussed in Section~\ref{s:regular}, for each $i$, construct by exhaustive search $E_i: \zo^{n_i} \times \zo^{n_i} \mapping \zo^{m_i}$ a $(\sigma/2, 2)$-regular function, where $m_i = i^2$.

We recall that this means that for all $k_1, k_2 \geq (\sigma/2) n_i$, for all $B_1 \subseteq \zo^{n_i}$ with $\lll A \rrr \geq 2^{k_1}$, for all $B_2 \subseteq \zo^{n_i}$ with $\lll B_2 \rrr \geq 2^{k_2}$, and for all $a \in \zo^{m_i}$,
\[
\lll E_i^{-1}(a) \cap B_1 \times B_2 \rrr \leq \frac{2}{2^{m_i}} \lll B_1 \times B_2 \rrr.
\]
We take $z_i = E_i (x_i, y_i)$.

Finally $z = z_1 z_2 \ldots z_i \ldots$.

}
}
\medskip

It is obvious that the above procedure is a truth-table reduction (\ie, it halts on all inputs).

In what follows we will assume that the two input sequences $x$ and $y$ have randomness rate $\tau$ and our goal is to show that the output $z$ has randomness rate $(1-\delta)$ for any $\delta > 0$.
\begin{lemma}
\label{l:block}
For any $\epsilon > 0$, for all  $i$ sufficiently large, $K(z_i \mid \bar{x}_{i-1} \bar{y}_{i-1}) \geq (1-\epsilon) \cdot m_i$.
\end{lemma}
$\proof$ 
Suppose $K(z_i \mid \bar{x}_{i-1} \bar{y}_{i-1}) < (1-\epsilon) \cdot m_i$. 

Let $A = \{z \in \zo^{m_i} \mid K(z  \mid \bar{x}_{i-1} \bar{y}_{i-1}) < (1-\epsilon) \cdot m_i\}$. We have $\lll A \rrr < 2^{(1-\epsilon)m_i}$.

Let $t_1, t_2, B_1, B_2$ be defined as follows:
\begin{itemize}
\item $t_1 = K(x_i \mid  \bar{x}_{i-1} \bar{y}_{i-1})$.  

Since $K(x_i \mid \bar{x}_{i-1}) > \sigma n_i$, and taking into account Lemma~\ref{l:conditional}, it follows that $t_1 > \sigma n_i - O(i) > (\sigma/2) n_i$, for all $i$ sufficiently large.

\item $t_2 = K(y_i \mid  \bar{x}_{i-1} \bar{y}_{i-1})$.

By the same argument as above, $t_2 > (\sigma/2) n_i$.

\item $B_1 = \{x \in \zo^{n_i} \mid K(x \mid  \bar{x}_{i-1} \bar{y}_{i-1}) \leq t_1 \}$.

\item $B_2 = \{y \in \zo^{n_i} \mid K(y \mid  \bar{x}_{i-1} \bar{y}_{i-1}) \leq t_2 \}$.

\end{itemize}

We have $\lll B_1 \rrr \leq 2^{t_1 + 1}$. Take $B_1'$ such that $\lll B_1' \rrr = 2^{t_1 + 1}$ and $B_1 \subseteq B_1'$. 

We have $\lll B_2 \rrr \leq 2^{t_2 + 1}$. Take $B_2'$ such that $\lll B_2' \rrr = 2^{t_2 + 1}$ and $B_2 \subseteq B_2'$. 

The bounds on $t_1$ and $t_2$ imply that $B_1'$ and $B_2'$ are large enough for $E_i$ to satisfy the regularity property on them. In other words, for any $a \in \zo^{m_i}$, 
\[
\lll E^{-1}_i (a)  \cap B_1' \times B_2' \rrr \leq \frac{2}{2^{m_i}} \lll B_1' \times B_2' \rrr.
\]
So,
\begin{equation*}
\begin{array}{ll}
\lll E^{-1}_i (A)  \cap B_1 \times B_2 \rrr & \leq \lll E^{-1}_i (A)  \cap B_1' \times B_2' \rrr  \\

& = \sum_{a \in A} \lll E^{-1}_i (a)  \cap B_1' \times B_2' \rrr \\

& \leq 2^{(1-\epsilon)m_i} \frac{2}{2^{m_i}} \lll B_1' \times B_2' \rrr \\

& \leq 2^{t_1 + t_2 - \epsilon m_i + 3}.
\end{array}
\end{equation*}
There is an algorithm that, given $(x_1, x_2, \ldots, x_{i-1})$, 
$(y_1, y_2, \ldots, y_{i-1})$, $(1-\epsilon)m_i$, $t_1$ and $t_2$,  enters an infinite loop during which it enumerates the elements of the set $E^{-1}_i (A)  \cap B_1 \times B_2$.  Therefore, the Kolmogorov complexity of any element of $E^{-1}_i (A)  \cap B_1 \times B_2$ is bounded by its rank in some fixed enumeration of this set, the binary encoding of the input (including the information needed to separate the different components),  plus a constant number of bits describing the enumeration procedure.

Formally, for every $(u,v) \in E^{-1}_i (A)  \cap B_1 \times B_2$,
\[
K(uv \mid \bar{x}_{i-1} \bar{y}_{i-1}) \leq t_1 + t_2 - \epsilon m_i + 2(\log(1-\epsilon)m_i + \log t_1 + \log t_2) + O(1) = t_1 + t_2 - \Omega(i^2).
\]
We took into account that $m_i = i^2$, $\log t_1 = O(i)$, and $\log t_2 = O(i)$.
In particular,
\[
K(x_i y_i \mid \bar{x}_{i-1} \bar{y}_{i-1}) \leq t_1 + t_2 - \Omega(i^2).
\]
On the other hand, by Lemma~\ref{l:combination},
\begin{equation*}
\begin{array}{ll}
K(x_i y_i \mid \bar{x}_{i-1} \bar{y}_{i-1} ) & \geq K(x_i  \mid \bar{x}_{i-1} \bar{y}_{i-1} ) + K(y_i \mid \bar{x}_{i-1} \bar{y}_{i-1}) - O(i) \\
& = t_1 + t_2 - O(i).
\end{array}
\end{equation*}
The last two inequations are in conflict, and thus we have reached a contradiction.~\qed
\begin{comment}
(PANIC: This is a crucial point in the proof, and the above is what I want to happen because of the independence of the two sources $x$ and $y$. Check it carefully

I need:

(1) $K(uv \mid w) \geq K(u \mid w) + K(v \mid wu) - O(\log|u| + \log|v|)$. This seems to be reasonable.
\smallskip

and
\smallskip

(2) $K(y_i \mid x_1 y_1 \ldots x_{i-1} y_{i-1} x_i ) \geq K(y_i \mid x_1 y_1 \ldots x_{i-1} y_{i-1}) - O(1)$.

I hope (2) follows from:

(2.1) $K^x(y_i \mid x_1 y_1 \ldots x_{i-1} y_{i-1}) \leq K(y_i \mid x_1 y_1 \ldots x_{i-1} y_{i-1} x_i ) - O(1)$. This seems reasonable.
\smallskip

and
\smallskip

(2.2) $K(y_i \mid x_1 y_1 \ldots x_{i-1} y_{i-1}) \leq K^x(y_i \mid x_1 y_1 \ldots x_{i-1} y_{i-1}) +O(1)$, which should somehow follow from the independence condition \ie, from  $K^x(y(1:n))\geq K(y(1:n)) - O(1)$, for almost every $n$.)
\end{comment}
%To have a contradiction, I need $\epsilon m_i > O(i)$. Therefore, we take $m_i = i^2$ .~\qed
\medskip

The following lemma concludes the proof of the main result.

\begin{lemma}
\label{l:final}
For any $\delta > 0$,  the sequence $z$ obtained by concatenating in order $z_1, z_2, \ldots$, has randomness rate at least $1- \delta$.
\end{lemma}

$\proof$ Take $\epsilon = \delta/4$.

By Lemma~\ref{l:block}, $K(z_i \mid \bar{x}_{i-1} \bar{y}_{i-1}) \geq (1-\epsilon) \cdot m_i$, for all $i$ sufficiently large.

This implies $K(z_i \mid z_1 \ldots z_{i-1}) > (1-\epsilon)m_i - O(1) > (1-2 \epsilon)m_i$ (because each $z_j$ can be effectively computed from $x_j$ and $y_j$).

By induction, it can be shown that $K(z_1 \ldots z_i) \geq (1-3\epsilon) (m_1 + \ldots + m_i)$. For the inductive step, we have
\begin{equation*}
\begin{array}{ll}
K(z_1 z_2 \ldots z_{i}) & \geq K(z_1 \ldots z_{i-1}) + K(z_i \mid z_1 \ldots z_{i-1}) - O(\log(m_1 + \ldots + m_{i-1}) + \log(m_i))\\

& \geq (1-3\epsilon)(m_1 + \ldots + m_{i-1}) +(1 - 2\epsilon)m_i - O(\log(m_1 + \ldots + m_i))\\

& > (1- 3\epsilon) (m_1 + \ldots + m_i).
\end{array}
\end{equation*}

Now consider some $z'$ which is between $z_1 \ldots z_{i-1}$ and $z_1 \ldots z_i$, \ie, for some strings $u$ and $v$, $z'= z_1 \ldots z_{i-1} u$ and $z_1 \ldots z_i = z'v$. Suppose $K(z') < (1-4\epsilon) |z'|$. 

Then $z_1 \ldots z_{i-1}$ can be reconstructed from: 

(a) the descriptor of $z'$, which takes $(1-4\epsilon)|z'| \leq (1-4\epsilon)(m_1 + \ldots +m_i)$ bits,

(b) the string $u$ which takes $|z'| - |z_1 \ldots z_{i-1}| \leq m_i$ bits

(c) $O(\log m_i)$ bits needed for separating $u$ from the descriptor of $z'$ and for describing the reconstruction procedure.

This implies that 
\begin{equation*}
\begin{array}{ll}
K(z_1 \ldots z_{i-1}) & \leq (1-4\epsilon)(m_1 + \ldots + m_i) + m_i +O(\log m_i)\\

& = (1-4\epsilon)(m_1 + \ldots + m_{i-1}) + (2-4\epsilon) m_i  + O(\log m_i) \\

&  = \big( 1 - 4\epsilon + (2-4\epsilon) \frac{m_i}{m_1 + \ldots + m_{i-1}} \big) \cdot(m_1 + \ldots +m_{i-1}) + O(\log m_i) \\

&<(1-3\epsilon)(m_1 + \ldots + m_{i-1}).
\end{array}
\end{equation*}

(The last inequality holds if $\frac{m_i}{m_1 + \ldots + m_{i-1}}$ goes to $0$, which is true for $m_i = i^2$.) This is a contradiction.

Thus we have proved that for every $n$ sufficiently large, $K(z(1:n)) > (1-\delta)n$.~\qed
\medskip

The main result can be stated in terms of constructive Hausdorff dimension, a notion introduced in measure theory. The constructive Hausdorff dimension of a sequence $x \in\{0,1\}^\infty$ turns out to be equal to $\lim \inf \frac{K(x(1:n))}{n}$(see~\cite{may:j:dimension-kol, rya:j:dimension, sta:j:dimension}). 
\begin{corollary}
For any $\tau > 0$, there is a truth-table reduction $f$ such that if $x \in\{0,1\}^\infty$ and $y\in\{0,1\}^\infty$ are independent and have constructive Hausdorff dimension at least $\tau$, then $f(x,y)$ has Hausdorff dimension $1$. Moreover, $f$ is uniform in the parameter $\tau$.
\end{corollary}
We next observe that Theorem~\ref{t:main} can be strengthened by relaxing the requirement regarding the independence of the  two input sequences. For a function $g: \nat \mapping \real^+$, we say that two sequences $x \in\{0,1\}^\infty$ and $y \in\{0,1\}^\infty$ have dependency $g$, if for all natural numbers $n$ and $m$,
\[
K(x(1:n)) + K(y(1:m)) - K(x(1:n)y(1:m)) \leq O(g(n) + g(m)).
\]
In Theorem~\ref{t:main}, the assumption is that the two input sequences have dependency $g(n) = \log n$.
%We say that a function $g: \N \mapping \real^+$ has \emph{moderate growth} if $\lim_{n \rightarrow %\infty} g(n)/(g(1) + g(2) + \ldots + g(n-1)) = 0$. 
Using essentially the same proof as the one that demonstrated Theorem~\ref{t:main}, one can obtain the following result.
\begin{theorem}
 For any $\tau > 0$, there exist $0 < \alpha < 1$ and  a truth-table reduction $f: \zo^\infty \times \zo^\infty \mapping \zo^\infty$  such that if $x \in\{0,1\}^\infty$ and $y \in\{0,1\}^\infty$ have dependency $n^{\alpha}$ and randomness rate $\tau$, then $f(x,y)$ has randomness rate $1-\delta$, for any positive $\delta$. Moreover, $f$ is uniform in the parameter $\tau$.
\end{theorem}
\begin{comment}
PROOF - SKETCH: In Section 4.2 all the O(i) become O(n^\alpha). To obtain the contradiction in Lemma 4.11~\ref{l:block}, we need to take $m_i = n^\beta = B^i$, (for $\beta > \alpha$) for some constant $B$ which can be taken to be arbitrarily close to 1 (but larger than 1). Construction of regular function everything is ok.  lemma 4.2 also works: lim m_i/(m_1 + m_2 + ... + m_{i-1}) is not zero but can be made an arbitrarily small constant as it is B-1.
\end{comment}
 In Theorem~\ref{t:main} it is required that the initial segments of $x$ and $y$ have Kolmogorov complexity at least $\tau \cdot n$, for a positive constant $\tau$. We do not know if it is possible to obtain a similar result for sequences with lower Kolmogorov complexity. However, using the same proof technique, it can be shown that if $x$ and $y$ have their initial segments with Kolmogorov complexity only $\Omega(\log n)$, then one can produce an infinite sequence $z$ that has very high Kolmogorov complexity for infinitely many of its prefixes. 
\begin{theorem}
For any $\delta > 0$, there exist a constant $C$ and a truth-table reduction $f: \zo^\infty \times \zo^\infty \mapping \zo^\infty$ with the following property: 

If the input sequences $x$ and $y$  are independent and satisfy $K(x(1:n)) > C \cdot \log n$ and $K(y(1:n)) > C \cdot \log n$, for every $n$, then the output $z = f(x,y)$ satisfies $K(z(1:n)) > (1- \delta) \cdot n$, for infinitely many $n$. Furthermore, there is an infinite computable set $S$, such that $K(z(1:n)) > (1- \delta) \cdot n$, for every $n \in S$.
\end{theorem}
\begin{comment}
PROOF (sketch): The main modifications are:

The lengths $n_i$ increase much faster. $n_i = A^{n_1 + \ldots + n_{i-1}}$, for some constant $A$.

The analogue of Lemma 4.7 will be with the slack term $- O(i)$ replaced by $-D\log n_i$, for some absolute constant $D$.

The sizes of the boxes in the probabilistic argument will be $n_i^C$ and $m_i = C' \log n_i$ for $C' < C$ so that the prob. argument works.

In Lemma 4.11, to obtain a contradiction we need to take $C'$ big enough (so $C$ has to be big enough).

In Lemma 4.12, we only get $K(z_1 \ldots z_i) \geq (1-3\epsilon) (m_1 + \ldots + m_i)$, but not the result for the intermediate $z'$ because we cannot get the ratio $m_i / (m_1 + \ldots + m_{i-1})$ to be small. 
~\qed.
\end{comment}
\section{Acknowledgments} The author thanks Ted Slaman for bringing to his attention the problem of extracting Kolmogorov complexity from infinite sequences, during the Dagstuhl seminar on Kolmogorov complexity in January 2006. The author is grateful to Cristian Calude for insightful discussions.

\newcommand{\etalchar}[1]{$^{#1}$}

%\bibliography{c:/book-text/theory}

%\bibliographystyle{alpha}
\end{document}